\documentclass{article}
\usepackage{cite}
\usepackage{amsmath,amsthm, amssymb,amsfonts}
\usepackage{graphicx}
\usepackage{textcomp}
\usepackage{xcolor}
\usepackage{comment}
\def\BibTeX{{\rm B\kern-.05em{\sc i\kern-.025em b}\kern-.08em
    T\kern-.1667em\lower.7ex\hbox{E}\kern-.125emX}}

\usepackage{algpseudocode}
\usepackage{algorithm}

\usepackage{authblk}

\newtheorem{theorem}{Theorem}

\algnewcommand\algorithmicforeach{\textbf{for each}}
\algdef{S}[FOR]{ForEach}[1]{\algorithmicforeach\ #1\ \algorithmicdo}

\begin{document}

\title{Synthesis Cost-Optimal Targeted Mutant Protein Libraries\\
}

\author[1]{Dimitris Papamichail}
\author[1]{Madeline Febinger}
\author[1]{Shm Almeda}
\author[2]{Georgios Papamichail}
\affil[1]{Department of Computer Science\\
\textit{The College of New Jersey}\\
Ewing, NJ, USA \\
papamicd@tcnj.edu}
\affil[2]{{School of Informatics}\\
\textit{New York College}\\
Athens, Greece \\}

\maketitle

\begin{abstract}
Protein variant libraries produced by site-directed mutagenesis are a useful tool utilized by protein engineers to explore variants with potentially improved properties, such as activity and stability. These libraries are commonly built by selecting residue positions and alternative beneficial mutations for each position. All possible combinations are then constructed and screened, by incorporating degenerate codons at mutation sites. These degenerate codons often encode additional unwanted amino acids or even STOP codons. Our study aims to take advantage of annealing based recombination of oligonucleotides during synthesis and utilize multiple degenerate codons per mutation site to produce targeted protein libraries devoid of unwanted variants. Toward this goal we created an algorithm to calculate the minimum number of degenerate codons necessary to specify any given amino acid set, and a dynamic programming method that uses this algorithm to optimally partition a DNA target sequence with degeneracies into overlapping oligonucleotides, such that the total cost of synthesis of the target mutant protein library is minimized. Computational experiments show that, for a modest increase in DNA synthesis costs, beneficial variant yields in produced mutant libraries are increased by orders of magnitude, an effect particularly pronounced in large combinatorial libraries. 
\end{abstract}


\section{Introduction}

Mutant libraries representing protein variants are increasingly used to optimize protein function. Protein Engineering involves screening mutant libraries for novel proteins that show enhanced expression levels, solubility, stability, or enzymatic activity. To reach such objectives, it is often advantageous to modify extant proteins and develop mutant variants with potentially improved properties \cite{Reetz2007, Parker2010}. However, there exists a massive space of potential mutations to consider.

Computational design of combinatorial libraries \cite{Voigt2002, Meyer2003, Pantazes2007, Treynor2007} provides a reasonable approach in the development of improved variants. Library-design strategies seek to experimentally evaluate a diverse but focused region of sequence space in order to improve the likelihood of finding a beneficial variant. 
Such an approach is based on the premise that prior knowledge can inform generalized predictions of protein properties, but may not be sufficient to specify individual, optimal variants. Libraries are particularly appropriate when the prior knowledge does not admit detailed, robust modeling of the desired properties, but when experimental techniques are available to rapidly assay a pool of variants.

The design of mutant protein libraries typically involves a manual process in which required sites for mutation are selected and ambiguous {\it degenerate} codons (those containing mixtures of nucleotides) are designed to introduce controlled variation in these positions. This is particularly useful in cases where definitive decisions regarding specific amino acid substitutions are non-obvious~\cite{Reetz2007}. The design of the protein variant library is complemented by use of synthesized degenerate oligonucleotides which enable annealing based recombination. Custom oligonucleotide overlaps enable the targeted introduction of crossovers at only desired positions, in turn enabling the desired level and type of diversity in a combinatorial library.

Traditional mutant protein library design methods involve the incorporation of a single degenerate codon (thereafter referred to as {\it decodon}) at each position where amino acid substitutions are considered. Decodons contain ambiguous ({\it degenerate}) bases, as shown in Table~\ref{tab:table1}. Degenerate bases are one letter codes are used to represent (i.e. code) sets of DNA bases.

\begin{table}[!htb]
    \centering
    \caption{Degenerate Bases and their codings}
    \begin{tabular}{c|c}
    \textbf{Degenerate Base} & \textbf{Actual Bases Coded}  \\
    N & A or C or G or T\\
    B & C or G or T\\
    D & A or G or T\\
    H & A or C or T\\
    V & A or C or G\\
    K & G or T\\
    M & A or C\\
    R & A or G\\
    S & C or G\\
    W & A or T\\
    Y & C or T\\
    \end{tabular}
    \label{tab:table1}
\end{table}

An online tool called CodonGenie~\cite{Swainston2017} was created to aid the effort of designing decodons that code for any provided set of amino acids. The CodonGenie tool ranks candidate decodons by specificity, attempting to minimize coding of undesired amino acids and/or STOP codons. Even so, when using a single decodon to code for a set of amino acids, it is often unavoidable to code for additional unwanted amino acids. Using an example from~\cite{Swainston2017}, when coding the non-polar residues A, F, G, I, L, M and V, CodonGenie picks decodon DBK ([AGT][CGT][GT]) as its top choice, which, in addition to the desired set, codes also for amino acids C, R, S, T, and W. In total, the decodon DBK codes for 26 DNA variants, 18 of which code for desired amino acids, and 8 DNA variants for undesired ones.

In our work we explore specifying a set of amino acids by using potentially multiple decodons. The usage of annealing based recombination of degenerate oligos containing such decodons can produce libraries on the productive portion of the space by eliminating unwanted mutations, therefore improving the yield of beneficial variants and the overall quality of the library. In turn, this method can significantly reduce labor costs assaying the pool of variants, at the expense of additional oligo synthesis, whose comparative cost is modest and continuously dropping. We further use the design of minimum cardinality decodon sets specifying any amino acid set (henceforth referred to as {\it AA-set}) to create an algorithm that, given a target protein mutant library, it designs oligos whose assembly generates the target library without any unwanted variants, while minimizing the total cost of DNA synthesis.

The remainder of the this paper is organized as follows. In section \ref{sec:defs} we provide definitions for describing our problem and proposed algorithms to solve it. In section \ref{sec:target} we explore an algorithm for provably finding the minimum number of decodons necessary to specify any given amino acid set. In section \ref{sec:small_pars} we present our main algorithm that aids the design of necessary sets of oligos to assemble our targeted protein variant library while minimizing the cost of synthesis. Experimental results in applying our algorithms toward building model variant libraries from the literature are discussed in section \ref{sec:results}. Finally, in section \ref{sec:conclusion} we summarize our findings and discuss future directions pursued in our lab.

\section{Definitions}\label{sec:defs}

For our purposes, a protein, peptide, or amino acid chain will be represented by a string on the 20-letter alphabet $\Sigma_{P} =$ \{A, C, D, E, F, G, H, I, K, L, M, N, P, Q, R, S, T, V, W, Y\}. A DNA sequence will comprise of bases
derived from the alphabet $\Sigma_{D} = \{A, C, G, T\}$. A {\it degenerate base} is a representation of multiple possible alternative nucleotides at a certain position. We
will use the International Union of Pure and Applied Chemistry (IUPAC) notation
to represent degenerate bases or {\it multi-bases}, DNA bases derived from a set of possible
alternatives. For example, the multi-base $R$ will indicate a Purine, a DNA base
that is either an A or a G. As such, the character 'R' in a DNA sequence will substitute for the regular expression $(A|G)$.

Each amino acid is encoded by one or more codons, triplets of DNA bases. Degenerate codons (thereafter referred to as {\it decodons}) are codons that can include multi-bases. For example, ARC is a decodon
representing both AAC or AGC. As such, ARC serves as a shorthand of the regular expression
$A(A|G)C$, or the equivalent $A[AG]C$ in common programming language notation. A decodon can code for more than one amino acid. For example, the decodon ARC codes for both Asparagine (AAC) and Serine (AGC).

A DNA sequence can be synthesized in the lab by joining overlapping DNA fragments called {\it oligonucleotides} or {\it oligos}.
Given an amino acid sequence $a$ of length $m = |a|$, it can be encoded by a DNA sequence
$d$ of length $n = |d| = 3 \times m$. Such a sequence can in principle be assembled by $k$ oligos of length $l_i$
each, $1 \le i \le k$, with each consecutive pair sharing an overlap of length $o_j$, $1 \le i \le k-1$, such that
$n = \sum_{i=1}^{k}{l_i} - \sum_{j=1}^{k-1}{o_j}$.

The input to our problem consists of an amino acid sequence of length $m$
and a list $p$ of $b$ positions $p_i, 1 \le i \le b$. For each given position, we are provided an amino
acid set $aa_i, 1 \le i \le b$ of desired amino acid substitutions. This input represents a protein variant
library $L$ with size $|L| = \prod_{i=1}^{b} |aa_i|$.

The desired output to our problem is a set of partially overlapping oligos that, once assembled, generate all $|L|$ mutant protein variants in the target library, and only those. In addition, the total number of DNA nucleotide bases in the produced oligos, necessary to assemble the target library, is minimized.


\section{Targeting protein libraries without undesired variants}\label{sec:target}

\begin{figure*}[!tbh]
\centering
	\includegraphics[width= 0.95\linewidth]{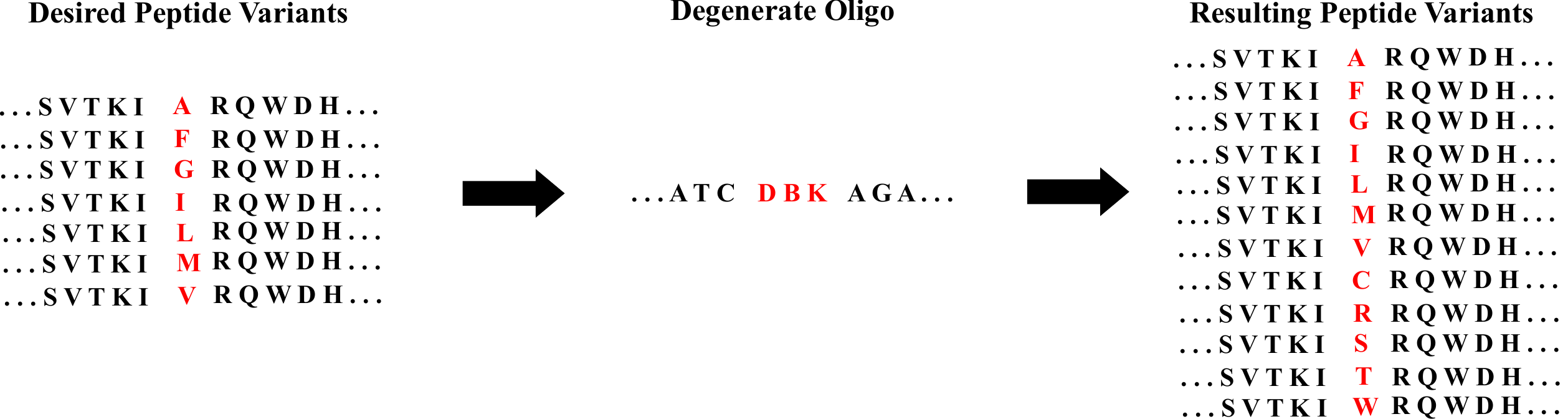}
    \caption{Single decodon use for encoding a variable amino acid position}
\label{fig:single_decodon}
\end{figure*}


Traditionally, protein variant libraries that vary amino acid residues at certain mutation sites are constructed by utilizing a single decodon at each variable position that can generate all targeted residues. Mutant protein variants with undesired amino acids at mutation sites can be avoided when synthesizing a targeted library. Instead of designing a single decodon to specify a given AA-set, we could design several decodons to specify the same set. Each of these decodons would code for a subset of the input AA-set, where the union of all these subsets would equal the input set. Then, for the creation of the protein variant library, we could synthesize multiple oligos, each incorporating a different individual decodon at the target mutant site.

\begin{figure*}[!thb]
\centering
	\includegraphics[width= 0.95\linewidth]{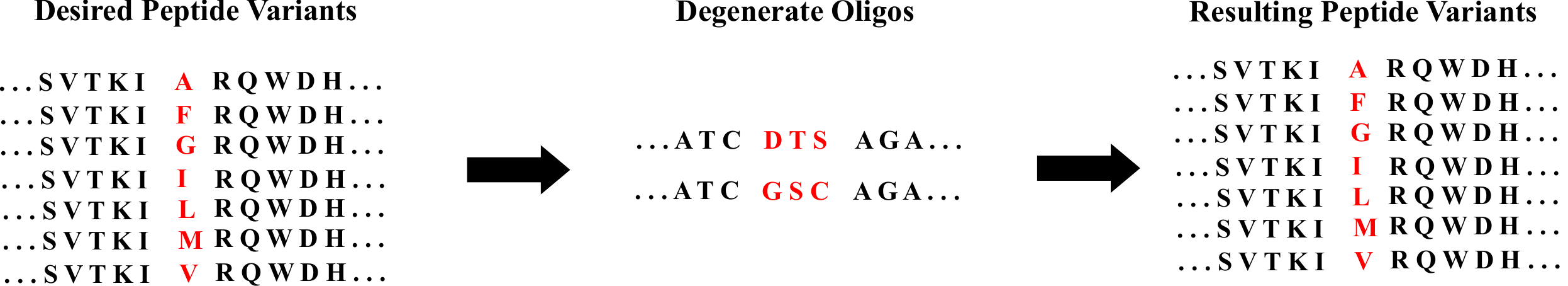}
    \caption{Double decodon use for encoding a variable amino acid position}
\label{fig:double_decodon}
\end{figure*} 

For example, suppose that at a mutant site of a given peptide we want to vary the residue from one of \{A, F, G, I, L, M, V\}, as shown in Figure \ref{fig:single_decodon}. The CodonGenie tool indicates in \cite{Swainston2017} that the single decodon DBK (regular expression $(A|G|T)(C|G|T)(G|T)$) can code for the given set, while minimizing the additional undesired amino acids coded, in this case the set \{C, R, S, T, W\}. If we wanted to eliminate these undesired amino acids, two decodons are necessary. One such decodon set is \{DTS, GSC\} (or $\{(A|G|T)T(C|G), G(C|G)C\}$, as shown in Figure \ref{fig:double_decodon}. Note that, even though we are using two decodons in two separate degenerate oligos instead of one, the number of actual oligos that will be incorporated in the target mutant library, $8$ in the latter case, is smaller than in the library using the single decodon at the mutation site of interest, which are $18$ in total.

Our aim is to minimize the amount of DNA we synthesize, which is directly proportional to the total cost of synthesis, while generating a targeted variant library with no undesired variants. Thus we seek to minimize the number of decodons at each variable amino acid position. The question becomes, what is the minimum number of decodons necessary to code any given amino acid set? Calculating the answer is the subject of the following subsection.

For the sake of simplicity, we will assume that the cost of each synthesized nucleotide base in our library is uniform, independent of whether that base is degenerate or not. In reality, DNA synthesis companies charge a slightly higher price for degenerate compared to regular bases. That additional cost can be amortized among all DNA bases ordered and included in the conservative price estimates that we use in our experimental results section.

\subsection{Optimal coding of amino acid subsets using degenerate codons}\label{sec:enum}

We have designed and implemented an algorithm that, given any set of amino acids, produces the minimum number of decodons necessary to code for exactly this set, i.e. without coding for extraneous amino acids or STOP codons. 
Decodons are triplets of the 15 nucleotide codes shown in table \ref{tab:table1}. As such there are $15^3 = 3,375$ decodons that can be assembled from this 15-letter alphabet of ambiguous codes.

The target input for our problem is an AA-set, a subset of all 20 possible amino acids. Since each amino acid may be included in the subset or not, the number of possible subsets is $2^{20} - 1$ (the empty set is excluded as an invalid input), i.e., there are 1,048,575 possible subsets of the 20 amino acids. 
Each of these sets can be represented by a 20-digit binary number, where a one at position n indicates that amino acid n is included in the set, and a zero indicates that it is absent. 

We initially explored calculating the minimum number of decodons needed to specify any AA-set by exhaustively considering all decodon combinations. We would start by figuring all AA-sets coded by a single decodon, then two decodons, etc. Unfortunately this method is practically intractable, even though our input size is constant, as the number of decodon combinations explored increases exponentially by a factor of $3375$ for each additional decodon considered. Similarly intractable is the direction of examining AA-sets and considering all their subset partitions.

A careful examination of our problem indicates that, once we know the answer for a given set $a$, we can potentially utilize this result to compute an answer for a proper superset $b$ of $a$, $b \supset a$, by combining the decodons specifying $a$ and those of any superset of $b\setminus a$. Owing to the associativity and commutativity of set union, any order that set unions are performed, as well as any way they are grouped, is not going to affect the resulting union when joining decodon sets, each specifying an individual amino acid subset.

Based on these observations, we designed the following algorithm: We start by computing all possible AA-sets that can be specified by a single decodon, which we will call 1-decodon AA-sets. Then we perform set unions of these AA-sets with themselves, to uncover all AA-sets specified by a minimum of $2$ degenerate codons, 2-decodon AA-sets. We then continue computing set unions between 1-decodon AA-sets and k-decodon AA-sets, with $2 \le k \le 19$, or until all $2^{20}-1$ sets are encountered.

Using our algorithm we calculated minimum cardinality decodon sets for all $1,048,575$ possible amino acid subsets. Our results indicate that $6$ decodons are always sufficient to code for any amino acid subset, where at most 4 decodons are sufficient to encode more than $90\%$ of all amino acid subsets. Our algorithm also produces an example of a decodon set of minimum cardinality for each AA-set computed.

The following is the pseudocode of our {\it MinDecodon} 
algorithm. AA-set list $L_i$ keeps track of all AA-sets 
specified by a minimum of $i$ decodons, and array $h$ 
tracks all AA-sets that have been encountered so far. AA-sets are 
represented as binary numbers, and set unions are performed as binary disjunctions. The function {\it Decodon\_to\_AA\_set} returns the AA-set specified by a given input decodon.

\medskip
\begin{algorithm}
\caption{The MinDecodon Algorithm}
\label{algo:min_decodon}
\begin{algorithmic}[1]
\State Initialize AA-set lists $L_{i}, 1 \le i \le 20$
\For{$1 \le i \le 2^{20}$}
\State $h[i] \xleftarrow{} 0$
\EndFor
\ForEach {decodon $d$}
\State aa\_set $\xleftarrow{}$ Decodon\_to\_AA\_set($d$)
\If{$h[d] = 0$}
\State $h[$aa\_set$] \xleftarrow{} 1$
\State add aa\_set to $L_1$
\EndIf
\EndFor
\medskip
\For{$1 \le rank \le 19$}
\ForEach{aa\_set1 in $L_1$}
\ForEach{aa\_set2 in $L_{rank}$}
\State combined\_aa\_set $\xleftarrow{}$ aa\_set1 $\cup$ aa\_set2
\If {h$[$combined\_aa\_set$] \ne 0$}
\State $h[$combined\_aa\_set$] \xleftarrow{} rank + 1$
\State add combined\_aa\_set to $L_{rank+1}$
\EndIf
\EndFor
\EndFor
\EndFor

\end{algorithmic}
\end{algorithm}
\medskip

The following theorem establishes that the MinDecodon algorithm always computes the minimum number of decodons necessary to specify any AA-set.

\begin{theorem}
The MinDecodon algorithm can correctly identify all AA-sets that require a minimum of $n$ decodons to exactly code for them.
\label{thm:decodon}
\end{theorem}
\begin{proof}
We will prove the statement by induction.

Base case: For $n = 1$, we need to show that the MinDecodon algorithm correctly identifies all AA-sets that can be specified by a single decodon. That is true, since the first step of the algorithm exhaustively processes all individual decodons and records the AA-sets that they specify.

Inductive hypothesis: We assume that the MinDecodon algorithm can correctly identify all AA-sets requiring at least $k$ or fewer decodons to specify them (strong hypothesis), where k is an arbitrary fixed integer with $1 \le k \le 20$.

Inductive step: We will prove that the statement is true for $n = k+1$. We do not have to argue about AA-sets that can be specified by $k$ or fewer decodons, since the inductive hypothesis assures the truth of the statement in these cases. Let us select an arbitrary AA-set $A_{any}$ that can be specified by $k+1$ decodons. We can list these decodons in any order as $d_1, d_2, \cdots, d_{k+1}$. Let $A_1$ be the AA-set specified by decodon $d_1$, $A_2$ the set specified by $d_2$, etc. Let $B = A_2 \cup A_3 \cup ... \cup A_{k+1}$. $A_1$ is an AA-set that is specified by a single decodon, where $B$ is a set specified by $k$ decodons. By the inductive hypothesis, both sets have been correctly identified by the MinDecodon algorithm. At the $k$th iteration, the MinDecodon algorithm considers the union of all sets specified by a single decodon and those specified by $k$ decodons. Therefore, AA-set $A_{any}$ will be constructed as the union of $A_1$ and $B$, thus being identified as a set that can be specified by $k+1$ decodons.
\end{proof}

\begin{figure}[!thb]
\centering
	\includegraphics[width= \linewidth]{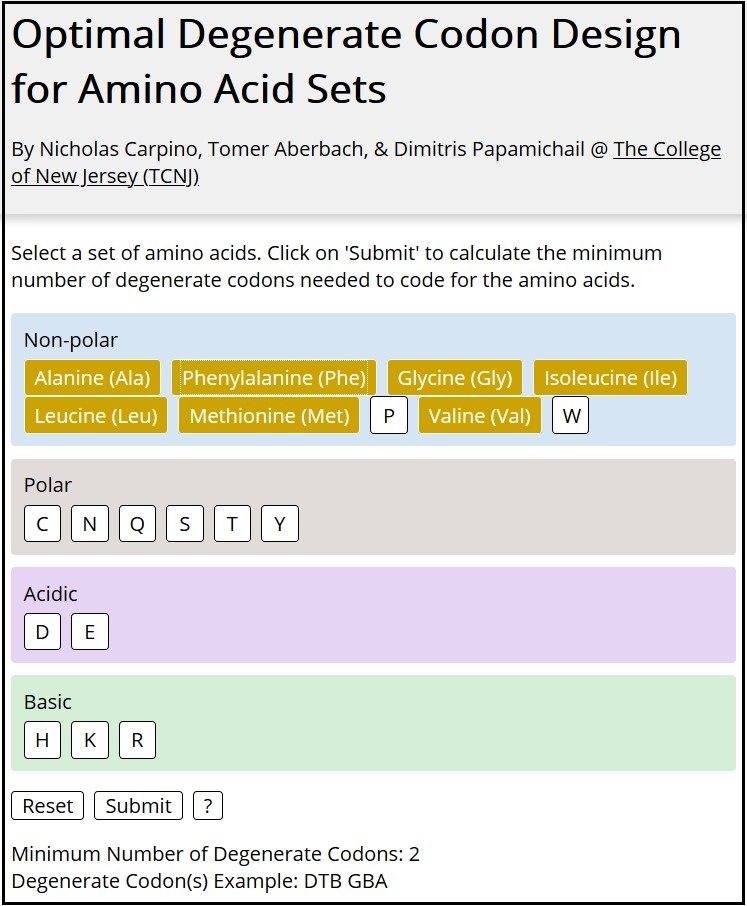}
    \caption{Calculating the minimum number of decodons necessary to encode the AA-set \{A, F, G, I, L, M, V\} using the Decodon Calculator web-tool}
\label{fig:results}
\end{figure} 

We also built a web tool called {\it Decodon Calculator} that allows researchers to view the minimum number of decodons needed to code for any input amino acid subset. 
Once a set of amino acids is selected and the Submit button is pressed, results are displayed at the bottom of the screen, as shown in Figure~\ref{fig:results}. In that particular example, we can observe that the non-polar residues A, F, G, I, L, M and V can be specified by the two decodons DTB and GBA, which code for 12 desirable DNA variants, in contrast to the 26 variants of the single best decodon generated by CodonGenie, 8 of which are undesired.

The Decodon Calculator can be accessed at 

http://algo.tcnj.edu/decodoncalc/.

\section{Optimal oligo design for synthesis cost minimization}\label{sec:small_pars}

In creating libraries of targeted protein variants, we enable substitutions of residues at pre-specified positions with alternatives drawn from AA-sets of beneficial variants, each corresponding to a mutation site. In this section we aim to optimize the combinatorial assembly of all such protein variants without any undesired residues at any position, while minimizing the total cost of synthesis. To achieve this goal, we limit the use of decodons at each mutation site to the exact minimal set that can code exactly for the corresponding AA-set, as calculated in the previous section by the algorithm MinDecodon.

Each protein in a mutant variant library is translated and transcribed from synthetic DNA, which is in turn assembled by joining multiple DNA oligos. It is this process that allows us to distribute degeneracies among different oligos and combinatorially combine the oligos to create libraries with large numbers of variants without synthesizing separate protein coding DNA sequences for each. Assembly methods such as the Gibson isothermal assembly \cite{Gibson} provide certain freedom for varying the length of the oligos and their overlaps, the latter usually ranging in length between 20-40 bases. By carefully selecting the breakpoints where the sequence is partitioned into oligos, we can reduce the total amount of DNA sequence that is required for the synthesis any given target mutant library.

\begin{figure}[!thb]
\centering
	\includegraphics[width=\linewidth]{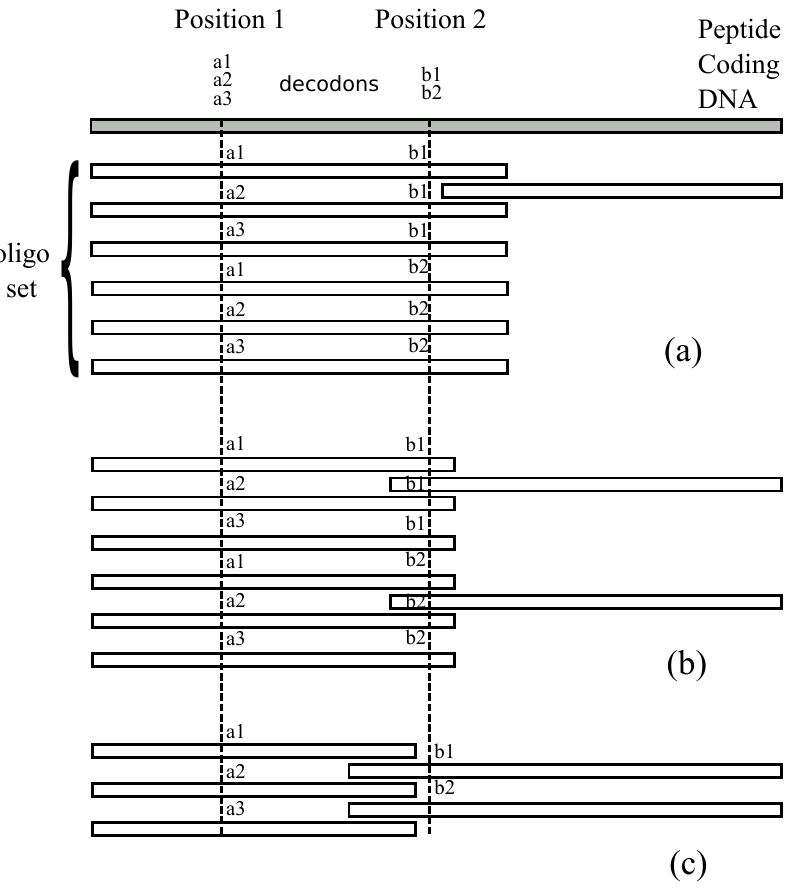}
    \caption{Examples of sequence to oligo partitioning and its effect on oligo synthesis cost.}
\label{fig:oligo_breaks}
\end{figure} 

As an example, let us consider three scenarios in breaking a target DNA sequence into two oligos. We assume that a target DNA sequence codes for a library of peptide variants where two sites are mutated, as depicted in Figure \ref{fig:oligo_breaks}. Each of the positions $1$ and $2$ are mutation sites that code for two provided AA-sets, $s1$ and $s2$. Let us assume that three decodons, $a1$, $a2$, and $a3$, are necessary to code for AA-set $s1$, where two decodons $b1$ and $b2$ are needed for set $s2$.

In Figure \ref{fig:oligo_breaks}(a) we depict a scenario where both mutation sites are placed in a single {\it oligo set}, i.e. a set of alternative oligos having the same start and end positions in the sequence. In this case, a total of $6$ oligos will need to be synthesized for the first oligo set, one for each combination of the $3$ decodons in position $1$ and the $2$ decodons in position $2$. With the addition of a single oligo without degeneracies representing the second oligo set, this design requires the synthesis of a total of $7$ oligos. In Figure \ref{fig:oligo_breaks}(b) we demonstrate what happens when a mutation site is placed inside an overlap. In such a case, decodons $b1$ and $b2$ need to be incorporated in both oligo sets, increasing the required number of total oligos to $8$. A best case scenario is shown in Figure \ref{fig:oligo_breaks}(c), where the mutation sites are placed in separate oligo sets. The first set now requires the synthesis of $3$ oligos, one for each decodon at position $1$, where the second set requires $2$ oligos, for a total of $5$ synthesized oligos.

Based on these observations, we aim to design an algorithm that, given as input
\begin{itemize}
    \item an amino acid sequence of length $m = n/3$, 
    \item a list of locations of mutation sites and number of decodons for each site, 
    \item and length ranges for oligos $(l_{min}, l_{max})$ and overlaps $(o_{min}, o_{max})$, with $0 < o_{min} \le o_{max} < l_{min} \le l_{max}$,
\end{itemize}
seeks to output a set of oligo set breakpoints, defined as pairs of start and end positions for each oligo set, such that, when the defined oligo sets are combinatorially assembled, they generate the targeted mutagenesis library at minimum synthesis cost.

Our algorithm uses dynamic programming to exhaustively consider all possible solutions to our problem, while storing partial optimal solutions for prefixes of the protein coding DNA sequence in a single-dimensional array of size $n$. The sufficiency of a linear partial solution space is based on the observation that, to compute the optimal cost of a final oligo of length $l_{final}$ sharing an overlap of length $o_{final}$ with a prefix of the DNA sequence ending at position $x = n - l_{final} + o_{final}$, only the cost of synthesis of that prefix is required as prior knowledge.

There are certain prefixes of any input sequence for which solutions cannot be computed. These include ending positions between 1 and $l_{min}$, between $l_{max}$ and $2 \times l_{min} - o_{max}$, etc. We do not need to treat these cases specially, since all positions in the partial solution array are initialized to $\infty$ and the first position to 0, with the dynamic programming process determining all unreachable positions during the algorithm execution.

The pseudocode of the {\it OligoBreak} algorithm is presented below.

\medskip
\begin{algorithm}
\caption{The OligoBreak Algorithm}
\label{algo:oligobreak}
\begin{algorithmic}[1]
\State $n \xleftarrow{}$ sequence length in DNA bases
\State $l_{min} \xleftarrow{} $ minimum oligo length
\State $l_{max} \xleftarrow{} $ maximum oligo length
\State $o_{min} \xleftarrow{} $ minimum overlap length
\State $o_{max} \xleftarrow{} $ maximum overlap length
\State Initialize all positions of cost array $c$ to $\infty$
\State $c[0] \xleftarrow{} 0$
\For{$l_{min} \le i < n$}
\For{$l_{min} \le j \le l_{max}$}
\For{$o_{min} \le k \le o_{max}$}
\State $start \xleftarrow{} i - j$
\If{$(start \ge 0) \wedge (c[start + k] \ne \infty)$ }
\State $o\_cost \xleftarrow{} calculate\_cost(start, i)$
\State $current\_cost \xleftarrow{} cost[start+k] + o\_cost$
\If{$current\_cost < cost[i]$}
\State $cost[i] \xleftarrow{} current\_cost$
\EndIf
\EndIf
\EndFor
\EndFor
\EndFor
\end{algorithmic}
\end{algorithm}
\medskip

We now proceed with proving the correctness of the OligoBreak algorithm using the following theorem:

\begin{theorem}
We will prove that the OligoPartition algorithm can optimally design oligos that assemble to form DNA sequences of length $n$, coding for a given targeted protein variant library, while minimizing the total number of DNA nucleotides synthesized.
\label{thm:oligopartition}
\end{theorem}
\begin{proof}
We will prove theorem \ref{thm:oligopartition} by induction.

Base case: The base case consists of all sequence lengths $k$ that can be represented by a single oligo, i.e. $l_{min} \le k \le l_{max}$. The optimality of their design follows by the principles of construction that were discussed in the previous section, where the number of decodons which specify any given AA-set is minimized.

Inductive hypothesis: We assume that the theorem statement holds true $\forall n: 0 \le n \le k$, where k is a fixed arbitrary integer.

Inductive step: We will prove that the statement is true for $n = k+1$. 
The main functionality of the algorithm depends on three nested loops, 
the outer loop iterating over the length of the sequence, the middle loop that iterates over the 
length range of an oligo, and the inner loop which iterates over the length range of an overlap. As 
such, during the $k+1$ iteration of the outer loop, the middle loop examines 
oligo sets of all possible lengths that end at the $k+1$ position, where the 
inner loop examines all possible overlap lengths. The cost of each of these possible oligo sets ending at position $k+1$ 
is calculated optimally, then added to the value of the cost array 
position based on the overlap being considered.

For the sake of contradiction, let us assume that the algorithm fails to
compute the minimum number of DNA bases needed to synthesize the
designed oligos ending at position $k+1$. Then there exists another
sequence of oligo sets with a better total cost. This sequence will end
with a final set that has a legal length $l_f$ and overlap $o_f$. Since
the algorithm examines all possible oligo and overlap lengths for the
final set and, for each case, determines the optimal oligo set based on
the principles of the decodon design of theorem \ref{thm:decodon}, this
final oligo set has been considered by the algorithm. Therefore, the
total cost of the oligo sets ending at position $k+1 - l_f + o_f$ must
be suboptimal (and by construction it has to be $l_f > o_f$). But that
contradicts our inductive hypothesis, that the cost array stores optimal
costs for all sequences of lengths $\le k$.
\end{proof}

\section{Experimental Results}\label{sec:results}

In this section we present comparative results from computational experiments that involve the creation of protein and protein segment variant libraries of interest using our oligo multiplexing method against ordering a single synthetic construct with degeneracies. The latter is designed using the CodonGenie tool to determine single decodons that specify given sets of amino acids at specific mutation sites. The computational experiments we performed are presented in order of increasing complexity and size of the protein variant library.

All computational experiments were performed utilizing parameter values as
described in Gibson et al. \cite{Gibson, Gibson2008, Gibson2009}. Individual oligonucleotides are
permitted to vary in length between $40$ and $90$ base pairs, while permissible overlaps between oligos ranged from $20$ to $40$ base pairs. It should be noted that these values are used only for reference and our algorithm can accept any reasonable value range for these parameters. For all cost calculations we will assume a cost of synthesis of \$$0.38$ per nucleotide, making the simplifying assumption that this cost includes possible degeneracies.

\subsection{Bacterial Type IV Pili}

Our first computational experiment involves a pilin-based $20$-mer peptide, which self-assembles into ordered nanofibres, as described in
\cite{Guterman2016}. This $20$-mer is a core peptide building block that is derived from the parent bacterial type IV pilus protein, a class of polymeric nanofibres that emerge from the surface of Gram-negative and Gram-positive bacteria and archaea, and are involved in diverse biological processes. The 20-mer sequence is "FTLIELLIPQFSUYRVKUYN".

Based on personal communication with Dr. Baker, a computational chemist at
The College of New Jersey, we compiled a list of desirable amino acid substitutions that would generate an interesting peptide library of $256$ variants for possible property enhancement of the $20$-mer peptide. This involved substitutions as follows:
\begin{itemize}
\item E5 $\rightarrow$ A5 or D5 or K5 or R5
\item P9 $\rightarrow$ A9 or G9
\item R15 $\rightarrow$ A15 or K15 or E15 or D15
\end{itemize}

The original amino acids at these positions were also included in the library variants. Each of these three sets of amino acids, $\{E, A, D, K, R\}$, $\{P, A, G\}$, and $\{R, A, K, E, D\}$, can be minimally encoded by 2 decodons each. The optimal library would only utilize one 60-base oligo for each variant; the resulting variant library would need the synthesis of a total of 480 bases and generate only the 108 desired variants. At \$0.38 per base, this library would cost \$182.40.
The same library, when utilizing a single decodon for each AA-set, as suggested by CodonGenie, would require the synthesis of a single 60-base long oligo at a cost of $22.80$, but create an additional {\it 468} undesired variants.

\subsection{Green Fluorescent Protein (GFP)}

In their work "Optimization of Combinatorial Mutagenesis" \cite{Parker}, Parker et al. developed an algorithm that selects optimal positions and sets of mutations to create a combinatorial mutagenesis library. Their
algorithm has the ability to create libraries that either utilize degenerate oligos or point mutagenesis. They tested their algorithm on the wild type 238-residue GFP from Aequorea victoria with the mutation S65T.

Figure 2 in Parker et al. \cite{Parker} displays examples of sets of 
mutations for different possible libraries, and ranks these libraries 
based on quality and novelty. As our target library we selected one with
a significant number of variants. For example, we used our algorithm
on the first library listed under degenerate oligos, which varies amino
acids at eight positions, $10 [EG]$, $53 [LV]$, $73 [AR]$, $124 [EK]$, $161 [IV]$, $162 [KR]$, $228 [GS]$. Our algorithm generates a variant library 
which requires the synthesis of a total of 969 nucleotides, coding for
the 256 variants. The oligos required for such a library would cost \$368.22 to
order. Instead, if we used a single decodon per position as 
recommended by CodonGenie, the library would require 903 bases, for a 
total DNA synthesis cost of \$343.14. However, the use of a single 
decodon per position results in coding 2 unwanted amino acids 
at position 73. Thus this latter library contains 512 total variants,
with 256 desired and 256 undesired variants. For an additional cost of 
only \$25, all of these
undesired variants can be avoided.

\subsection{Antiapoptotic B-cell lymphoma-extra large (Bcl-xL)}

Our third experiment was performed on structure based re-design of the binding specificity of anti-apoptotic protein Bcl-xL \cite{Chen2013}; a particularly interesting dataset due to the moderate number of varied positions but comparatively large sets of amino acid variants at each position. The resulting target variant libraries have sizes in the order of $10^6$ to $10^7$.

B-cell lymphoma-extra large (Bcl-xL) is a member of the B-cell lymphoma protein family (Bcl-2). The Bcl-2 family is anti-apoptotic, meaning it prevents cells from naturally dying. Researching the Bcl-2 family helps us understand cell death, which has potential applications to cancer therapeutics. Bcl-2 proteins interact with many partners, such as BH3 motifs. The peptides used to bond to these BH3 motifs are called BH3 peptides. Bim and Bad are BH3 peptides. The aim of this study was to redesign the anti-apoptotic protein Bcl-xL to prevent it from strongly interacting with Bim BH3, yet still keeping a tight binding to a BH3 peptide derived from Bad.

\begin{table}[!htb]
    \centering
    \caption{Bcl-xL anti-apoptotic protein target variants from Chen et al.}
    \begin{tabular}{c|c|c|c}
\textbf{Position} & \textbf{Amino acid set} & \textbf{Position} & \textbf{Amino acid set} \\
    F97 & AFGILMV & E96 & ADEFGHIKLMNQRSTVY\\
    Y101 & AFGILMTVY & Y101 & HY\\
    A104 & AFGILMSTVY & A104 & AFMW\\
    L108 & AFGILMV & L108 & LRTV\\
    L112 & AFGILMV & Q111 & ADEFGHIKLMNQRSTVY\\
    V126 & AFGILMV & S122 & ADEFGHIKLMNQRSTVY\\
    E129 & AEITV & Q125 & ADEFGHIKLMNQRSTVY\\
    L130 & AFGILMV & V126 & AV\\
    A142 & AGSTV & E129 & ETV\\
         &       & L130 & LI\\
         &       & F146 & AFGILMV\\
         &       & Y195 & FY\\
    \end{tabular}
    \label{tab:table2}
\end{table}

In their study, Chen et al. initially selected 9 sites of Bcl-xL at which to vary the amino acid residues, 
as depicted in Table 1 in their paper.  We created our first dataset 
by encoding all of the underlined non-disruptive amino acids from the
“amino acids modeled” column of that table, which are shown in the first
two columns of Table \ref{tab:table2}. For our target protein we used the 
211-residue chain A of PDB Bcl-xL. On this dataset, our algorithm
generates oligos whose synthesis requires ordering a total of 2952 nucleotide 
bases, which, at \$0.38 a base, would cost \$1,121.76 to synthesize.
Using CodonGenie, the total synthesis requirement goes down to 801 bases, 
at a cost of \$304.38. This latter library encodes a total of $7.1 \times 10^7$ 
variants, where only $3.1 \times 10^6$ of those variants are targeted. 
Only 1 out of 23 variants in that library is desired. We also 
experimented with a second targeted variant library from the same study,
specified in Table 2 of their paper. Once again, the dataset targets the
underlined non-disruptive amino acids in the “amino acids modeled” 
column. These variant positions and corresponding AA-sets are 
shown in column 3 and 4 of table \ref{tab:table2}. Our algorithm 
generates a library requiring oligos totaling 2604 nucleotides, with a 
projected cost of \$989.52. Using CodonGenie a total of 801 nucleotides
need to be synthesized, costing \$304.38. However, the latter library 
contains a total of $1.5 \times 10^{11}$ variants, where only $2.3 \times 10^9$ are 
desired, or 1 out of every 65 variants.

\section{Conclusion}\label{sec:conclusion}

In this study we examined the problem of designing oligos for synthesis of targeted protein variant libraries without unwanted mutations. Toward that goal, we also implemented an algorithm to determine the minimum number of decodons necessary to specify any given amino acid set. Using that result, we built an optimal dynamic programming algorithm to generate a set of oligos that, when assembled, create targeted protein variants and only those, while minimizing the cost of DNA synthesis. Compared to the traditional `one decodon per mutated site' approach, our oligo sets incur additional synthesis costs, but often increase the yield of useful variants in our libraries by orders of magnitude.

There are several future directions to be explored. Many DNA assembly techniques require melting temperatures among complementary overlapping oligo regions to fall within a certain range to maintain assembly efficiency. We are currently working on our next generation of oligo design algorithms that provide the option to utilize temperature ranges instead of overlap lengths when figuring optimal oligo breakpoints. 

DNA synthesis companies often place a limit on the number of degenerate bases that are introduced on each oligo ordered. This restriction, combined with the exponential growth of the number of required oligos when numerous mutation sites are located in close proximity, create a need for balancing the amount of degeneracy allowed with the quality of the produced library. We plan for next iterations of our algorithms and tools to incorporate constraints on the number of degenerate bases per oligo and the total cost of the library. The latter may be satisfied by reducing the number of decodons at certain sites that contribute less toward variant diversity, thus limiting the impact on reduced library quality. 

\section*{Acknowledgements}

We would like to thank Dr. Joseph Baker of the Chemistry Department at the College of New Jersey for valuable discussions and providing an example application for testing our algorithms.

This work has been supported in part by NSF Grant CCF-1418874 and The College of New Jersey Mentored Undergraduate Summer Experience (MUSE) program. The authors also acknowledge use of the ELSA high performance computing cluster at The College of New Jersey for conducting the research reported in this paper. This cluster is funded in part by the National Science Foundation under grant numbers OAC-1826915 and OAC-1828163.


\bibliographystyle{plain}
\bibliography{bibliography}

\end{document}